\documentclass[aps,prb,twocolumn,superscriptaddress]{revtex4} 

\usepackage[dvips]{graphicx}
\usepackage{bm}

\usepackage{color}
\usepackage{latexsym} 
\usepackage{amsmath}
\usepackage{amssymb}

\newcommand{\ui}{{\rm i}} 
\newcommand{\bmr}{{\bm r}}

\newcommand{\bms}{{\bm s}}
\newcommand{\bmS}{{\bm S}}

\newcommand{\bmq}{{\bm q}}

\newcommand{\bmk}{{\bm k}}

\newcommand{\bmp}{{\bm p}}
\newcommand{\kB}{k_{\rm B}} 

\newcommand{\bmm}{{\bm m}}
\newcommand{\bmx}{{\bm x}}
\newcommand{\bmy}{{\bm y}}

\newcommand{\bmH}{{\bm H}}
\newcommand{\bmh}{{\bm h}}
\newcommand{\bmE}{{\bm E}}
\newcommand{\bmJ}{{\bm J}}

\begin{document}
 
\title{
  Enhanced dc spin pumping into a fluctuating ferromagnet near $T_{{\rm C}}$ 
}
 
\author{Yuichi Ohnuma}
\email{y-ohnuma@imr.tohoku.ac.jp}
\affiliation{Institute for Materials Research, Tohoku University, Sendai 980-8577, Japan}
\affiliation{Advanced Science Research Center, Japan Atomic Energy Agency, Tokai 319-1195, Japan} 

\author{Hiroto Adachi}
\affiliation{Advanced Science Research Center, Japan Atomic Energy Agency, Tokai 319-1195, Japan}
\affiliation{CREST, Japan Science and Technology Agency, Sanbancho, Tokyo 102-0075, Japan}

\author{Eiji Saitoh}
\affiliation{Institute for Materials Research, Tohoku University, Sendai 980-8577, Japan}
\affiliation{Advanced Science Research Center, Japan Atomic Energy Agency, Tokai 319-1195, Japan}
\affiliation{CREST, Japan Science and Technology Agency, Sanbancho, Tokyo 102-0075, Japan}
\affiliation{WPI Research Center, Advanced Institute for Material Research, Tohoku University, Sendai 980-8577, Japan}

\author{Sadamichi Maekawa}
\affiliation{Advanced Science Research Center, Japan Atomic Energy Agency, Tokai 319-1195, Japan}
\affiliation{CREST, Japan Science and Technology Agency, Sanbancho, Tokyo 102-0075, Japan}

\date{\today}

\begin{abstract} 
A linear-response formulation of the dc spin pumping, i.e., a spin injection from a precessing ferromagnet into an adjacent spin sink, is developed in view of describing many-body effects caused by spin fluctuations in the spin sink. It is shown that when an itinerant ferromagnet near $T_{\rm C}$ is used as the spin sink, the spin pumping is largely with its Curie temperature located close to room temperature increased owing to the fluctuation enhancement of the spin conductance across the precessing ferromagnet/spin sink interface. As an example, the enhanced spin pumping from yttrium iron garnet into nickel palladium alloy ($T_{\rm C} \simeq 20$K) is analyzed by means of a self-consistent renormalization scheme, and it is predicted that the enhancement can be as large as tenfold.
\end{abstract} 


\maketitle 

\section{Introduction~\label{Sec:Intro}} 

There has been a growing demand for an efficient method of generating a spin current because it is a key quantity in driving the functionality of spintronic devices~\cite{Maekawa-text02}. In the early days, an idea of electrical spin current injection from a metallic ferromagnet into nonmagnetic metals or semiconductors was theoretically proposed~\cite{Aronov76}, and later on, it was successfully demonstrated in experiments~\cite{Johnson85}. Although such a technique is by now recognized as a standard method for the spin current injection~\cite{Takahashi08}, the method suffers from a problem called impedance mismatch, which means that a huge reduction in the spin injection efficiency appears when there is a large difference in resistivity between the ferromagnetic spin current injector and the spin current sink~\cite{Schmidt00,Rashba00}. Moreover, such an electrical spin injection method is available only when {\it both} the spin current injector and the spin current sink are electrically conducting. 

Recently a completely different type of spin injection method, termed spin pumping~\cite{Tserkovnyak02}, has attracted much attention as an alternative and efficient way for the spin injection~\cite{Saitoh06,Ando08-1,Kajiwara10,Patra12,Hahn13,Rojas13}. In this method, nonequilibrium dynamics of magnetization in a ferromagnet injector is driven by ferromagnetic resonance (FMR), and the precessing magnetization ``pumps'' spins into an adjacent spin sink by transferring spin angular momentum through the $s$-$d$ exchange interaction at the interface. The FMR-driven spin pumping has an advantage that it is unaccompanied by any charge transfer across the spin injector/spin sink interface, such that it is free from the impedance mismatch problem and thus available even when the injector is an insulating magnet~\cite{Kajiwara10}. Because of this versatility, the spin pumping in a variety of systems is now a subject of intensive research~\cite{Harii07,Ando08-2,Kajiwara09,Mosendz10,Ando10,Yoshino10,Nakayama10,Sandweg11,Iguchi11,Yoshino11,Heinrich11,Czeschka11,Rezende11,Ando11,Dazhi12,Ando12,Qiu12,Castel12,Shikoh13,Du13,Ando13,Chen13}. Furthermore, a spin injection from permalloy (Py) into GaAs by means of the spin pumping, which would otherwise suffer from the impedance mismatch problem, was successfully demonstrated~\cite{Ando11}. 

Originally, the FMR-driven spin pumping is formulated~\cite{Tserkovnyak02} in a close analogy to a theory of adiabatic charge pumping in mesoscopic systems~\cite{Brouwer98}. The efficiency of the spin pumping is then characterized by a quantity called {\it spin mixing conductance} $g^{\uparrow\downarrow}$, the value of which may be calculated by the Landauer-B\"{u}ttiker approach combined with first-principles calculation~\cite{Tserkovnyak05}. However the physical meaning of $g^{\uparrow\downarrow}$ and its microscopic origin are obscure in the existing literature~\cite{Tserkovnyak02}, and moreover, there is no knowledge at present on how to take account of many-body effects in the spin pumping theory. 

By contrast, the linear-response formalism is a powerful theoretical framework to describe nonequilibrium phenomena near thermal equilibrium. In particular, it is amenable to the language of the magnetism community, and more importantly, it has an advantage that it can easily deal with many-body effects when combined with a field-theoretical approach~\cite{Mahan-text}. In the context of spintronics, the linear-response approach has greatly contributed to the progress in a thermal version of the spin pumping, termed the spin Seebeck effect~\cite{Uchida08,Uchida10,Jaworski10,Rafa13,Adachi13,Ohnuma13}. It was not until the advent of the linear-response formulation of the spin Seebeck effect~\cite{Adachi11} that a description of the phonon-drag process~\cite{Adachi10}, which is now recognized as one of the principal mechanisms of the spin Seebeck effect~\cite{Uchida11}, was made possible. Therefore, it is quite natural to develop a linear-response formulation of the spin pumping in order to describe many-body effects. 

In this paper, we develop a linear-response theory of the FMR-driven dc spin pumping by using field-theoretical methods~\cite{Rammer86} in order to clarify the role of many-body effects. We investigate intriguing effects of critical spin fluctuations on the spin pumping, and show that when a metallic ferromagnet near the Curie temperature $T_{\rm C}$ is used as the spin sink, the spin pumping is largely enhanced owing to the fluctuation enhancement of the spin conductance across the spin injector/spin sink interface. Central to the above argument is the fact that the interface spin conductance (conventionally denoted as $g^{\uparrow \downarrow}$) is proportional to the imaginary part of the dynamical spin susceptibility of the spin sink, ${\rm Im} \chi_\bmk (\omega)$, which is known to be largely enhanced near $T_{\rm C}$~\cite{Hohenberg77}. This suggests that the interface spin conductance is effectively enhanced if the spin sink is made of an itinerant ferromagnet close to $T_{\rm C}$, and that the resultant spin pumping attains a large enhancement. This argument is justified in this work by a microscopic analysis which is supplemented by the self-consistent renormalization (SCR) theory of spin fluctuations~\cite{Moriya-text,Lonzarich85,Lonzarich89}. 

The plan of this paper is as follows. 
In the next section, we introduce a model Hamiltonian to describe the dc spin pumping. In Sec.~\ref{Sec:Form}, we present a linear-response formulation of the FMR-driven spin pumping that allows us to investigate many-body effects on the spin pumping. Next, in Sec.~\ref{Sec:Numer}, we apply the linear-response formalism to the dc spin pumping into a fluctuating itinerant ferromagnet near $T_{\rm C}$. For illustration, we analyze the spin pumping into a weak itinerant ferromagnet (NiPd alloy) from an insulating magnet (e.g., yttrium iron garnet) by using the SCR theory~\cite{Moriya-text,Lonzarich85, Lonzarich89}, and demonstrate that the spin pumping is largely enhanced close to the Curie temperature of the spin sink. The enhancement can be detected experimentally by observing either the additional Gilbert damping~\cite{Mizukami02} or the pumped spin current~\cite{Saitoh06}. Note that this enhanced spin pumping should be distinguished from the fluctuation effects on the spin Hall angle~\cite{Wei12,Gu12}. In Refs.~\onlinecite{Wei12} and \onlinecite{Gu12}, the anomaly in the inverse spin Hall effect at $T_{\rm C}$ due to skew scattering in NiPd alloy has been studied (see the inset of Fig.~\ref{fig5_Ohnuma} below). In this paper, we examine the fluctuation enhancement of the spin pumping (i.e., the dashed curve in the inset of Fig.~\ref{fig5_Ohnuma}). In Sec.~\ref{Sec:Conclusion} we summarize and discuss our result. In Appendix~\ref{Sec:Append03}, we briefly discuss the opposite case in which the spin injector is made of a fluctuating ferromagnet near $T_{\rm C}$ whereas the spin sink is a nonmagnetic metal without critical spin fluctuations.

\section{Model \label{Sec:Model}} 
\begin{figure}[t] 
  \begin{center}
    \includegraphics[width=6cm]{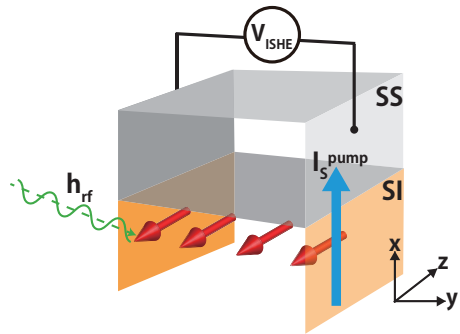}
  \end{center}
\caption{ 
(Color online) Schematic view of a system considered in the present paper for the FMR-driven spin pumping. A bilayer of a spin injector (SI) and a spin sink (SS) is placed in an external static magnetic field $\bmH_0$. The wavy line (green) represents an external ac magnetic field $\bmh_{\rm rf}$ which induces precessional motion of localized spins in the SI. 
}
\label{fig1_Ohnuma}
\end{figure}

The system for observing the FMR-driven spin pumping is a bilayer composed of a spin injector (SI) with precessing spins and an adjacent spin sink (SS), as shown in Fig.~\ref{fig1_Ohnuma}. While the SI can be either a ferromagnetic metal or a ferromagnetic insulator since the spin pumping is not accompanied by any charge transfer across the SI/SS interface, we consider here the case of an insulating SI to simplify the argument. As for the SS, a {\it nonmagnetic} metal, most typically Pt, is commonly used because Pt shows a relatively large inverse spin Hall effect that is necessary to electrically detect the pumped spin current~\cite{Saitoh06}. Although a use of Pt as the SS looks most promising, an itinerant ferromagnet such as NiPd alloy is also known to possess a moderate strength of the inverse spin Hall effect~\cite{Wei12,Gu12}, and hence it can be used as the SS in the spin pumping experiments. We are particularly interested in such a situation in which the SS is made of a weak itinerant ferromagnet having a relatively low $T_{\rm C}$ and possessing a sizable inverse spin Hall effect, e.g., a bilayer system composed of NiPd alloy/yttrium iron garnet. 

We begin with the following Hamiltonian:
\begin{equation}
  {\cal H} = {\cal H}_{\rm SS}+ {\cal H}_{\rm SI}+ {\cal H}_{\rm sd}+{\cal H}_{\rm rf}, 
  \label{H_total01}
\end{equation}
where the first term~\cite{Kawabata74}, 
\begin{eqnarray}
  {\cal H}_{\rm SS} &=& 
  \sum_{\bmp}  \epsilon_{\bmp} c^{\dag}_{\bmp} c_{\bmp'} 
  + U \sum_{i} n_{i\uparrow} n_{i\downarrow} \nonumber \\
  &+& 
  \sum_{\bmp,\bmp'} c^\dag_{\bmp} V_{\bmp-\bmp'} \Big[ 
    1  + \ui \eta_{\rm so} {\bm \sigma} \cdot (\bmp \times \bmp') \Big] 
  c_{\bmp'}, 
  \label{Eq:H_SS01} 
\end{eqnarray}
is the Hamiltonian describing the SS. Because we are interested in the case where the SS is a weak itinerant ferromagnet, we use a model of an itinerant ferromagnet with local electron-electron interaction~\cite{Kawabata74}. Here, $c^\dag_\bmp=(c^\dag_{\bmp,\uparrow},c^\dag_{\bmp,\downarrow})$ is the electron creation operator for spin projection $\uparrow$ and $\downarrow$, $\epsilon_\bmp$ is the kinetic energy of electrons, $U$ is the on-site Coulomb repulsion, and $n_{i,\sigma}= c^\dag_{i,\sigma} c_{i,\sigma}$ is the spin-projected charge density at a position $\bmr_i$, where $c_{i,\sigma}= N_{\rm SS}^{-1/2} \sum_{\bmp,\sigma} c_\bmp e^{\ui \bmp \cdot \bmr_i}$ with $N_{\rm SS}$ being the number of lattice sites in the SS. In addition to the kinetic and Coulomb terms describing a clean weak itinerant ferromagnet, we take account of impurity effects given by the Fourier transform, $V_{\bmp-\bmp'}$, of the impurity potential $V_{\rm imp} \sum_{\bmr_{\rm imp} \in \text{impurities}} \delta(\bmr- \bmr_{\rm imp})$ with $\eta_{\rm so}$ measuring the strength of the spin-orbit interaction~\cite{Takahashi08}. 
The second term, 
\begin{eqnarray}
  {\cal H}_{\rm SI} &=& 
  -J_{\rm ex} \sum_{\langle i,j \rangle \in {\rm SI}} 
  \bmS_i\cdot \bmS_j + \gamma \hbar \sum_{i \in {\rm SI}} H_0 S^z_{i}, 
  \label{Eq:H_SI01} 
\end{eqnarray} 
describes the SI, where $J_{\rm ex}$ is the nearest-neighbor exchange integral, $\bmS_i$ the spin operator at a position $\bmr_i$, $\gamma$ the gyromagnetic ratio, and $H_0$ the static magnetic field in the $z$ direction. The third term in Eq.~(\ref{H_total01}), 
\begin{eqnarray}
  {\cal H}_{\rm sd} &=& 
 J_{\rm sd}
 \sum_{i \in \text{SI/SS}\mathchar`-\text{interface}} 
 {\bm s}_i \cdot \bmS_{i}, 
  \label{Eq:H_sd01} 
\end{eqnarray} 
describes the interaction between the SI and the SS~\cite{Takahashi10}, where $J_{\rm sd}$ is the $s$-$d$ exchange interaction at the SI/SS interface, and $\bms_i= c^\dag_i {\bm \sigma} c_i$ with Pauli matrices ${\bm \sigma}$ is the itinerant spin density operator in the SS. 
Finally the last term in Eq.~(\ref{H_total01}), 
\begin{eqnarray}
  {\cal H}_{\rm rf} &=& 
  \gamma \hbar \bmh_{\rm rf} \cdot \Big( \sum_{i \in {\rm SI}}  \bmS_{i} \Big), 
  \label{Eq:H_rf01} 
\end{eqnarray} 
describes the effect on the SI of a circular polarized oscillating magnetic field 
$  \bmh_{\rm rf}(t) = h_{\rm rf} \cos (\Omega_{\rm rf} t) {\bm \hat{\bmx}}
  - h_{\rm rf} \sin (\Omega_{\rm rf} t) {\bm \hat{\bmy}} ,$ 
which is approximated to be spatially uniform since the wavelength of the oscillating field is longer than the sample dimension. 

Because we are interested in the low-energy excitation of localized spins in the SI that is driven by the oscillating field of GHz frequency ($\Omega_{\rm rf} \sim$ GHz), we use the spin-wave approximation. Introducing magnon variables $b^\dag_i$ and $b_i$ by means of the linear Holstein-Primakoff transformation 
\begin{eqnarray}
S^{x}_{i} + iS^{y}_{i} &=& \sqrt{2 S_0} b_i, \\
S^{x}_{i} - iS^{y}_{i} &=& \sqrt{2 S_0} b^\dag_i, 
\end{eqnarray}
Eq.~(\ref{Eq:H_SI01}) is diagonalized to be 
\begin{eqnarray}
  {\cal H}_{\rm SI} &=& \hbar \sum_\bmq 
  \omega_\bmq b^\dag_\bmq b_\bmq, 
  \label{Eq:S_SI02}
\end{eqnarray}
where $S_0= |\bmS_i| $ and $\hbar\omega_\bmq = 2 J_{\rm ex} z_0 S_0 (1-\gamma_\bmq)+ \hbar \gamma H_0 $ with $\gamma_\bmq= z_0^{-1} \sum_{\bm \delta} e^{\ui \bmq \cdot {\bm \delta}}$ being defined by the sum over $z_0$ nearest neighbors. Similarly, Eqs.~(\ref{Eq:H_sd01}) and (\ref{Eq:H_rf01}) become 
\begin{eqnarray}
  {\cal H}_{\rm sd} &=& \sqrt{\frac{2S_0}{N_{\rm SI}N_{\rm SS}}}
  \sum_{\bmk,\bmq} {\cal J}_{\rm sd}(\bmk,\bmq)
  \big[ s^+_\bmk b^\dag_\bmq + s^-_\bmk b_\bmq \big], 
  \label{Eq:H_sd02} \\
  {\cal H}_{\rm rf} &=& 
  \gamma \hbar \sqrt{2 S_0} \Big[ h^+_{\rm rf} b^\dag_{\bmq={\bm 0}} e^{-\ui \Omega_{\rm rf} t}+ 
    h^-_{\rm rf} b_{\bmq={\bm 0}} e^{\ui \Omega_{\rm rf} t} \Big], 
  \label{Eq:H_rf02} 
\end{eqnarray}
where 
$\bms_\bmk= \sum_\bmp c^\dag_{\bmp+\bmk} {\bm \sigma} c_\bmp$, $h^\pm_{\rm rf}= h^x_{\rm rf} \pm \ui h^y_{\rm rf}$, 
${\cal J}_{\rm sd}(\bmk,\bmq)= \sum_{i \in {\rm SI}/{\rm SS}} J_{\rm sd}e^{\ui (\bmk -\bmq)\cdot \bmr_i}$, and  $N_{\rm SI}$ is the number of lattice sites in the SI. We use Eqs.~(\ref{Eq:H_SS01}), (\ref{Eq:S_SI02}), (\ref{Eq:H_sd02}), and (\ref{Eq:H_rf02}) in the following analysis.

\section{Linear-Response Formulation of dc Spin Pumping \label{Sec:Form}} 

\begin{figure}[t] 
  \begin{center}
    \includegraphics[width=5cm]{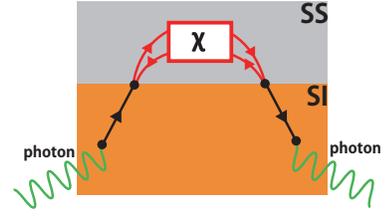}
  \end{center}
\caption{ 
(Color online) Feynman diagram representing the process of the dc spin pumping. The red and blue solid lines represent conduction-electron Green's function and uniform-mode magnon Green's function, respectively. The transverse susceptibility of itinerant spin density, $\chi_\bmk (\omega)$, is defined as a propagator of particle-hole pairs. The green wavy line represents external ac magnetic field $\bmh_{\rm rf}$. 
}
\label{fig2_Ohnuma}
\end{figure}

In this section, we present a linear-response formulation of the FMR-driven spin pumping by using field-theoretical methods~\cite{Rammer86}. The main purpose of our approach is to provide a theoretical framework to take account of many-body effects on the spin pumping. Therefore, we consider a situation where the SS acts as a perfect spin sink towards the conduction-electron spin current, by assuming that the thickness of the SS is comparable to the conduction-electron spin diffusion length of the SS (but much shorter than the magnon diffusion length of the SS). The FMR-driven spin pumping manifests itself in an appearance of both the pumped spin current~\cite{Saitoh06,Brataas02} into the SS and the additional Gilbert damping~\cite{Tserkovnyak02,Mizukami02} in the precessing SI. 

We first calculate the spin current pumped into the SS. Since we focus on the dc spin pumping, the pumped spin current $I_s^{\rm pump}$ has a polarization along the axis of magnetization precession in the SI which we take as the $z$ direction. The pumped spin current $I_s^{\rm pump}$ can be calculated as the rate of change of the itinerant spin density: 
\begin{eqnarray}
I_s^{\rm pump}&=&\sum_{i \in {\rm SS}} \langle \partial_t s_i^z \rangle, 
\end{eqnarray}
where $\langle \cdots \rangle$ denotes the statistical average. 
Note that the definition of the pumped spin current is similar to that of the tunneling current through a junction~\cite{Cohen62}, such that the time derivative of the itinerant spin density does not vanish even in the steady state and the spin current $I^{\rm pump}_s$ thus defined correctly describes the spin current pumped from the SI into the SS (see Eq.~(11) in Ref.~\onlinecite{Cohen62}). Note also that although a spin current in the form of magnon current may be pumped as well in the case of a ferromagnetic SS, the pumped magnon current is canceled by the backflow magnon current because we assume that the magnon diffusion length is much larger than the thickness of the SS.

Using $\bms_\bmk$ defined below Eq.~(\ref{Eq:H_rf02}), the pumped spin current can be expressed as $I_s^{\rm pump}= \sqrt{N_{\rm SS}} \langle \partial_t s^z_{\bmk_0 \to {\bm 0}} \rangle$. The right-hand side can be evaluated using the Heisenberg equation of motion for $s^z_{\bmk_0}$, giving 
\begin{eqnarray} 
  \partial_{t} s^{z}_{\bmk_0 \to {\bm 0}} 
  &=&   \frac{\ui}{\hbar} \sum_{\bmk, \bmq}
    \frac{\sqrt{2S}J_{\rm sd}(\bmk,\bmq)}{\sqrt{N_{\rm SI}N_{\rm SS}} } 
  b_\bmq s_{\bmk}^{-}+{\rm h.c.}, 
  \label{Eq:HeisenEOM}
\end{eqnarray}
where $s^\pm_\bmk= \frac{1}{2}(s^x_\bmk \pm \ui s^y_\bmk)$. Taking the statistical average of the above quantity, the right-hand side can be represented as 
\begin{eqnarray}
  I_s^{\rm pump} &=& -\frac{2\sqrt{2 S_{0}}}{\sqrt{N_{\rm SS}N_{\rm SI}}\hbar} 
  {\rm Re} \sum_{\bmk, \bmq} 
  J_{\rm sd}(\bmk,\bmq) C^{<}_{\bmk,\bmq}(t,t') \Big|_{t'\to t}, 
\label{Eq:Is_total01} 
\end{eqnarray}
where $C^{<}_{\bmk,\bmq}(t,t') = -\ui \langle b_{\bmq}(t') s_{\bmk}^{-}(t) \rangle$ is the interface Green's function defined by magnon operator $b_{\bmq}$ and the itinerant spin-density operator $s^-_\bmk$.  

In evaluating the right-hand side of Eq.~(\ref{Eq:Is_total01}), we adopt a diagrammatic approach with perturbation expansion in term of the external oscillating magnetic field $h_{\rm rf}$ and the $s$-$d$ interaction $J_{\rm sd}$ at the interface. The dc spin pumping process, which is proportional to the external microwave power and has a Lorentzian form, is given by the diagram shown in Fig.~\ref{fig2_Ohnuma}. Using the standard rules of evaluating the contour-ordered Green's function presented in Appendix~\ref{Sec:Append01}, the interface Green's function $C^<_{\bmk,\bmq}(t,t')$ is calculated to be 
\begin{eqnarray}
   C^<_{\bmk,\bmq}(t,t')
   &=& \ui \frac{J_{\rm sd}(\bmk,\bmq)}{\hbar} 
   \sqrt{\frac{S_0^{3}N_{\rm SI}}{2 N_{\rm SS}}} 
   (\gamma h_{\rm rf})^2   \delta_{\bmq,{\bm 0}}   \nonumber \\
   &\times&
   \int_{t_{1}, t_{2}, t_{3}} \chi^{R}_{\bmk}(t,t_{1}) 
   G^{(0)A}_{0}(t_{2},t_{1}) \nonumber \\ 
   && \qquad \qquad \times G^{(0)R}_{0}(t', t_{3}) e^{i\Omega_{\rm rf}(t_{2}-t_{3})}, 
   \label{Eq:C-func01}
\end{eqnarray}
where we introduced the shorthand notation $\int_{t_1,t_2,t_3}= \int_{-\infty}^\infty dt_1 dt_2 dt_3$. In the above equation, $G^{(0)R}_{0}(t_{3},t') = \ui\theta(t'-t_{3}) \langle [b_{\bmq={\bm 0}}(t_{3}), b_{\bmq={\bm 0}}^{+}(t')] \rangle_0$ and $G^{(0)A}_{0}(t_{3},t') = \ui\theta(t'-t_{3}) \langle [b_{\bmq={\bm 0}}(t_{3}), b_{\bmq={\bm 0}}^{+}(t')] \rangle_0$ are respectively the retarded and advanced parts of the {\it bare} Green's function of the uniform-mode magnon in the SI, whereas $\chi^{R}_{\bmk}(t, t_{1}) = \ui\theta(t-t_{1}) \langle [s^-_{\bmk}(t), s_{-\bmk}^{+}(t_{1})] \rangle$ is the retarded part of the transverse spin susceptibility in the SS that includes the interaction effects. 

Because each Green's function appearing in Eq.~(\ref{Eq:C-func01}) depends only on the difference of two time labels, it is advantageous to work in frequency space. Introducing the Fourier transform $\chi_{\bmk}(t_1,t_2)= \int_{-\infty}^{\infty} \frac{d \omega}{ 2 \pi} \chi_{\bmk}(\omega) e^{-i\omega (t_1-t_2)}  $ and $G^{(0)}_{0}(t_1,t_2)= \int_{-\infty}^{\infty} \frac{d \omega}{ 2 \pi} G^{(0)}_{0}(\omega) e^{-i\omega (t_1-t_2)}  $, Eq.~(\ref{Eq:C-func01}) becomes 
\begin{eqnarray} 
  C^<_{\bmk,\bmq}(t,t') &=& \ui \frac{J_{\rm sd}(\bmk,\bmq)}{\hbar} \sqrt{\frac{S_0^3 N_{\rm SI}}{2 N_{\rm SS}}} (\gamma h_{\rm rf})^2  \delta_{\bmq,{\bm 0}} \nonumber \\
   && \times \chi^{R}_{\bmk}(-\Omega_{\rm rf}) |G^{(0)R}_{0}(\Omega_{\rm rf})|^2 e^{i\Omega_{\rm rf}(t-t')} ,
\label{Eq:C-func02}
\end{eqnarray}
where we have used the general relation $G_0^{(0)R}(\omega)= [G_0^{(0)A}(\omega)]^{*}$. In the above equation, because we do not consider any anomalies occurring in the SI, the uniform-magnon Green's function is given in its bare form 
$G_0^{(0)R} (\omega) = 1/(\omega - \gamma H_0 + \ui \alpha_0 \omega)$ 
with $\alpha_0$ being the bare Gilbert damping. On the other hand, as schematically depicted in Fig.~\ref{fig2_Ohnuma}, the transverse spin susceptibility of the SS, $\chi^R_{\bmk}(\omega)$, includes the many-body effects. In the next section, the self-energy corrections caused by magnetic critical fluctuations in the SS is analyzed by means of a self-consistent renormalization scheme. 

Substituting Eq.~(\ref{Eq:C-func02}) into Eq.~(\ref{Eq:Is_total01}) and assuming the diffuse scattering of magnons at the SI/SS interface, we finally obtain the pumped spin current as 
\begin{eqnarray}
  I_s^{\rm pump} &=& 
  g_s  \frac{\Omega_{\rm rf} (\gamma h_{\rm rf})^2}
  {(\Omega_{\rm rf}-\gamma H_0)^2+ (\alpha_0 \Omega_{\rm rf})^2}, 
  \label{Eq:Is-total02b} 
\end{eqnarray}  
where $g_s$ represents the spin conductance across the SI/SS interface and is defined by 
\begin{eqnarray}
  {g}_{s} &=& \frac{2 J_{\rm sd}^2 S^2_0 N_{\rm int}}{\hbar^2 N_{\rm SS}}
  \sum_{\bmk} \frac{1}{\Omega_{\rm rf}}{\rm Im} \chi_\bmk^R(\Omega_{\rm rf}) 
  \label{Eq:Gs01}
\end{eqnarray}
with $N_{\rm int}$ being the number of localized spin $\bmS_i$ at the SI/SS interface. Note that the spin conductance is proportional to the momentum sum of the imaginary part of the dynamical {\it transverse} spin susceptibility ${\rm Im} \chi_\bmk^R(\omega)$, and that we have used the relation ${\rm Im} \chi_\bmk^R(-\Omega_{\rm rf}) = -{\rm Im} \chi_\bmk^R(\Omega_{\rm rf})$ to arrive at Eq.~(\ref{Eq:Gs01}).

\begin{figure}[t] 
  \begin{center} 
    \includegraphics[width=5cm]{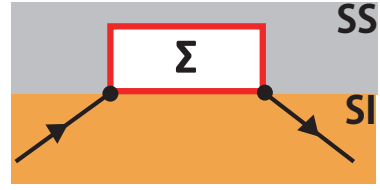}
  \end{center}
\caption{ 
(Color online) Diagrammatic representation of renormalized Green's function of uniform-mode magnon, ${G}_0^R(\omega)$, and the corresponding self-energy, $\Sigma^R_0(\omega)$. The black line with arrow means the bare magnon propagator ${G}_0^{(0)R}(\omega)$.} 
\label{fig3_Ohnuma}
\end{figure}

We next investigate the additional Gilbert damping caused by the dc spin pumping. Information on the damping of magnons at FMR is encoded in the imaginary part of the self-energy of uniform-mode magnon Green's function at the resonance frequency, $\Sigma^R_0(\omega=\Omega_{\rm rf})$, defined by the following Dyson's equation: 
\begin{eqnarray}
  \big[ {G}^R_0(\omega) \big]^{-1} &=& 
  \big[G^{(0)R}_0(\omega)\big]^{-1} - \Sigma^R_0(\omega), 
  \label{Eq:SPG-G001} 
\end{eqnarray}
where ${G}_0^R(\omega)$ and $G^{(0)R}_0(\omega)$ are respectively the renormalized and bare Green's functions of the uniform-mode magnon. Recalling that the imaginary part of $\Sigma^R_0(\omega)$ gives the damping rate of uniform-mode magnon and that in general the imaginary part of self-energy in a Bose system is proportional to its frequency $\omega$ at small $\omega$~\cite{AGD}, we obtain the relationship between the self-energy and additional Gilbert damping constant $\delta \alpha$ as 
\begin{eqnarray}
  \delta \alpha 
  &=& 
  - \frac{1}{\Omega_{\rm rf}}{\rm Im} \Sigma^R_{0} (\Omega_{\rm rf}) . 
  \label{Eq:Dalpha01} 
\end{eqnarray}

The renormalized Green's function ${G}_0^R(\omega)$ and self-enegy $\Sigma^R_0(\omega)$ associated with the dc spin pumping is diagrammatically given in Fig.~\ref{fig3_Ohnuma}. 
Comparing Dyson's equation (\ref{Eq:SPG-G001}) with Fig.~\ref{fig3_Ohnuma} and Fig.~\ref{fig2_Ohnuma}, we identify 
\begin{eqnarray}
  \Sigma^R_0 (\omega)   &=& -\frac{2 J^{2}_{\rm sd} S_0 N_{\rm int}}{\hbar^{2} N_{\rm SI}} \frac{1}{N_{\rm SS}}
  \sum_\bmk {\rm Im} \chi^{R}_\bmk(\omega). 
\label{Eq:SPG-G02}
\end{eqnarray}
Using the relation Eq.~(\ref{Eq:Dalpha01}), we obtain the additional Gilbert damping 
\begin{eqnarray}
  \delta \alpha &=& 
  \frac{1}{S_0 N_{\rm SI}} g_s, 
  \label{Eq:Dalpha02}
\end{eqnarray}
where $g_s$ is given in Eq.~(\ref{Eq:Gs01}). Equations~(\ref{Eq:Is-total02b}), (\ref{Eq:Gs01}), and (\ref{Eq:Dalpha02}) are the main results of this section. 

Before ending this section, it is instructive to discuss the relationship between the present formalism and that given in Ref.~\onlinecite{Tserkovnyak02}. In Ref.~\onlinecite{Tserkovnyak02}, the pumped dc spin current with $z$-axis polarization and the additional Gilbert damping are given by 
\begin{eqnarray}
  I^{\rm pump}_s &=& \frac{g^{\uparrow \downarrow}}{4 \pi} 
  \langle [\bmm \times \partial_t \bmm]^z \rangle, \label{Eq:pump01}\\
  \delta \alpha &=& \frac{\gamma \hbar }{4 \pi M_s {\cal V}} g^{\uparrow \downarrow}, 
  \label{Eq:pump02}
\end{eqnarray}
where $g^{\uparrow \downarrow}$ is the so-called spin mixing conductance, $\bmm$ is the magnetization direction vector, and $M_s$ and ${\cal V}$ are respectively the saturation magnetization and the volume of the ferromagnet. The above equations mean that the pumped spin current and the additional Gilbert damping are intimately related through 
\begin{eqnarray}
  I^{\rm pump}_s &=& \delta \alpha \frac{\gamma \hbar}{M_s {\cal V}} 
  \langle [\bmm \times \partial_t \bmm]^z \rangle. 
  \label{Eq:Is-Dalpha01} 
\end{eqnarray}
Using $\gamma \hbar /(M_s {\cal V}) = 1/(S_0 N_{\rm SI})$ and the expression 
\begin{eqnarray}
  \langle [\bmm \times \partial_{t}\bmm]^z \rangle 
  &=& \frac{\Omega_{\rm rf} (\gamma h_{\rm rf})^{2}}{(\Omega_{\rm rf}-\gamma H_0)^{2}+(\alpha_{0}\Omega_{\rm rf})^{2}}, 
  \label{Eq:SPG-SP01}
\end{eqnarray}
which applies in a region where the Landau-Lifshitz-Gilbert equation is valid, the consistency between our formalism and that given in Ref.~\onlinecite{Tserkovnyak02} [Eq.~(\ref{Eq:Is-total02b}) $\Leftrightarrow$ Eq.~(\ref{Eq:pump01}); and Eq.~(\ref{Eq:Dalpha02}) $\Leftrightarrow$ Eq.~(\ref{Eq:pump02})] can be confirmed with the identification
\begin{eqnarray}
  \frac{g^{\uparrow \downarrow}}{4 \pi} &=& g_s. 
\end{eqnarray}

\section{dc Spin Pumping into Fluctuating Ferromagnets near the Curie temperature\label{Sec:Numer}}  

In this section, we apply the formalism developed in the previous section to the dc spin pumping into a fluctuating ferromagnet near $T_{\rm C}$, and show that the resultant spin pumping is largely enhanced owing to the fluctuation enhancement of the interface spin conductance $g_s$. In the previous section we have shown, by deriving Eq.~(\ref{Eq:Is-total02b}), that the pumped spin current is determined by the following two factors: (i) the interface spin conductance $g_s$ which reflects information on the transverse susceptibility $\chi^R_\bmk(\omega)$ of the SS [see Eq.~(\ref{Eq:Gs01})], and (ii) the Lorentzian factor, which is equivalent to the magnetization damping torque $\langle [\bmm \times \partial_t \bmm]^z \rangle$ in the SI [see Eq.~(\ref{Eq:SPG-SP01})]. Because the imaginary part of $\chi^R_\bmk(\omega)$ is known to be enhanced near its $T_{\rm C}$ owing to the critical spin fluctuations~\cite{Hohenberg77,Moriya-text,Lonzarich85,Lonzarich89}, we can expect a fluctuation enhancement of $g_s$ and thus an enhanced spin pumping when the SS is made of an itinerant ferromagnet near $T_{\rm C}$. 

Let us first analyze the critical spin fluctuation effects on the transverse susceptibility $\chi^R_\bmk(\omega)$ by means of the SCR theory~\cite{Moriya-text,Lonzarich85}. In the following calculation, it is convenient to introduce the dimensionless transverse susceptibility 
\begin{eqnarray}
\widetilde{\chi}^{R}_\bmk (\omega) \equiv {{\chi}^{R}_\bmk (\omega)}/{\chi_{\rm P}}, 
\end{eqnarray}
where $\chi_{\rm P}$ is the Pauli paramagnetic susceptibility, and normalize length by the lattice spacing $d_0$. In the low frequency and long wavelength limit, the {\it bare} transverse susceptibility of the SS can be parametrized as 
\begin{eqnarray}
  \widetilde{\chi}^{(0)R}_\bmk (\omega) &=& 
  \frac{1}
       {\delta^{(0)} + c^{(0)} k^2 - \ui {\omega}/{\gamma^{(0)}_k}}, 
       \label{Eq:chi01}
\end{eqnarray}
where $c^{(0)}$ is the bare stiffness and $\gamma^{(0)}_k$ the bare damping rate of the spin fluctuations. The bare mass $\delta^{(0)}$ is given by $\delta^{(0)} = a^{(0)}+ b^{(0)} ({m}^{(0)})^2,$ where $a^{(0)} = A^{(0)}(T- T^{(0)}_{\rm C})/T^{(0)}_{\rm C}$ with a slope $A^{(0)}$ measures the distance from the transition temperature in the mean-field approximation $T^{(0)}_{\rm C}$, and $b^{(0)}$ is the bare mode-coupling constant. Here, the magnitude of a dimensionless magnetization ${m}^{(0)}$ is determined by the equation of state~\cite{Lonzarich85}: 
\begin{eqnarray}
  a^{(0)} {m}^{(0)} + b^{(0)} ( {m}^{(0)} )^3 
  &=& \widetilde{H}_{0}, 
  \label{Eq:EOS01}
\end{eqnarray}
where $\widetilde{H}_{0}= H_0/h_0$ is the dimensionless uniform external magnetic field normalized by $h_0=\gamma \hbar/(2\chi_{\rm P} v_0)$ with $v_0=d_0^3$ being the cell volume. 

If we apply the mean-field approximation to the Hamiltonian~(\ref{Eq:H_SS01}), we have~\cite{Moriya-text} $a^{(0)} \approx 1- U N(0)$ with $N(0)$ being the density of states of electrons at the Fermi energy. In a similar way, we have 
$b^{(0)}= (U^2/3!) \int \frac{d^3p}{(2 \pi)^3} [-\frac{d^3}{d \varepsilon_\bmp^3} f(\varepsilon_p)]$ with $f(\varepsilon_p)$ being the Fermi distribution function, and 
$c^{(0)}= (U^2/12) \int \frac{d^3p}{(2 \pi)^3} [\frac{d^3}{d \varepsilon_\bmp^3} f(\varepsilon_p) v^2_\bmp+  3 \frac{d^2}{d \varepsilon_\bmp^2} M_\bmp]$ with ${\bm v}_\bmp= d \varepsilon_\bmp/d \bmp$ and $M_\bmp= (1/2) \sum_{j=x,y,z} d^2 \varepsilon_\bmp/d p_j^2$. The bare damping rate in the presence of spin-orbit interaction is given by $\gamma^{(0)}_\bmk = D k^2 + \tau^{-1}_{\rm sf}$, where $D$ and $\tau_{\rm sf}$ are the spin diffusion coefficient and spin-flip relaxation time, respectively~\cite{Fulde68}. 

It is instructive to transform Eq.~(\ref{Eq:chi01}) into the form
\begin{eqnarray}
  \widetilde{\chi}^{(0)R}_\bmk (\omega) &=& 
  \frac{ \widetilde{\chi}^{(0)}_0 }{1+ \big( \xi^{(0)} k \big)^2 
    - \ui {\omega}/{\Gamma^{(0)}_k}}, 
  \label{Eq:chi02}
\end{eqnarray}
where $\widetilde{\chi}^{(0)}_0 = 1/\delta^{(0)}$ is the dimensionless uniform susceptibility, $\xi^{(0)}= \sqrt{c^{(0)}/\delta^{(0)} }$ is the effective correlation length, and $\Gamma^{(0)}_k= \gamma^{(0)}_k \delta^{(0)}$ is the effective damping rate. 
From this expression we see that, in the limit of vanishing external field ($\widetilde{H}_{0}=0$), the uniform susceptibility diverges as $\widetilde{\chi}^{(0)}_0 = (T-T^{(0)}_{\rm C})^{-1}$, and thus the divergent correlation length appears in $\xi^{(0)} \propto (T-T^{(0)}_{\rm C})^{-1/2}$ and the critical slowing down manifests itself in $\Gamma^{(0)}_k \propto (T-T^{(0)}_{\rm C})$. 

\begin{figure}[t] 
  \begin{center} 
    \includegraphics[width=5cm]{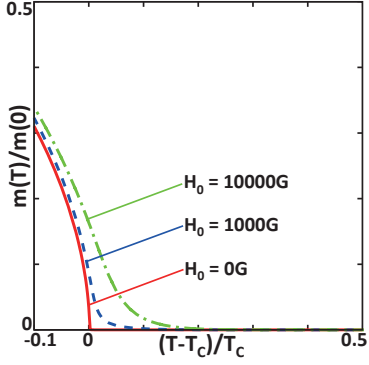}
  \end{center}
  \caption{ 
(Color online) Magnetization ${m}$ as a function of reduced temperature $(T-T_{\rm C})/T_{\rm C}$ for $H_0=0$G (solid line), $H_0=1000$G (dashed line), and $H_0=10000$G (dash-dotted line). The data is normalized by its value at $T=0$. 
  }
\label{fig4_Ohnuma}
\end{figure}

The SCR theory tells us how the bare transverse susceptibility $\chi^{(0)R}_\bmk (\omega)$ is modified into the renormalized transverse susceptibility $\chi^{R}_\bmk (\omega)$ due to the mode-mode coupling effect of magnetic critical fluctuations. In the dimensionless form, it is expressed as 
\begin{eqnarray}
  1/\widetilde{\chi}^{R}_\bmk (\omega) &=& 1/\widetilde{\chi}^{(0)R}_\bmk (\omega)
  + \Lambda, 
  \label{Eq:SCR01}
\end{eqnarray}
where the mode-coupling term $\Lambda$ is given by 
\begin{eqnarray}
  \Lambda = \frac{3 b}{N_{\rm SS}} \sum_\bmk \int_{-\infty}^\infty \frac{d \omega}{2 \pi}
  \coth \left( \frac{\hbar \omega}{ 2 \kB T }\right) {\rm Im} 
  \widetilde{\chi}^{R}_\bmk (\omega). 
  \label{Eq:SCR02}
\end{eqnarray}
The renormalized transverse susceptibility is assumed to have the following form: 
\begin{eqnarray}
  \widetilde{\chi}^{R}_\bmk (\omega) &=& 
  \frac{1}
       {\delta+ c k^2 - \ui {\omega}/{\gamma_k}}, 
       \label{Eq:chi03}
\end{eqnarray}
where $c$ is the renormalized stiffness and ${\gamma_k}$ is the renormalized damping rate. The renormalized mass $\delta$ is given by $\delta = a + b {m}^2,$ where $a \propto (T-T_{\rm C})/T_{\rm C}$ measures the distance from the renormalized Curie temperature $T_{\rm C}$, and $b$ is the renormalized mode coupling constant. Here, the magnitude of the magnetization ${m}$ is determined by Eq.~(\ref{Eq:EOS01}) with $a^{(0)}$ and $b^{(0)}$ being replaced by $a$ and $b$. Because the essential renormalization effect appears through the coefficient $a$ (and thus $\delta$), we set in the following $b = b^{(0)}$, $c = c^{(0)}$, and $\gamma_k= \gamma^{(0)}_k$ as is customarily done~\cite{Lonzarich85}. If we adopt the representation similar to Eq.~(\ref{Eq:chi02}), we obtain 
\begin{eqnarray}
  \widetilde{\chi}^{R}_\bmk (\omega) &=& 
  \frac{\widetilde{\chi}_0}
       {1 + \big( \xi k \big)^2 - \ui {\omega}/{\Gamma_k}}, 
\end{eqnarray}
where $\widetilde{\chi}_0= 1/\delta$, $\xi= \sqrt{c/\delta}$, and $\Gamma_k= \gamma_k \delta$. 

We calculate the renormalized mass $a$ self-consistently using Eqs.~(\ref{Eq:SCR01}) and (\ref{Eq:SCR02}) combined with the equations of state (\ref{Eq:EOS01}). In Fig.~\ref{fig4_Ohnuma}, we plot the magnetization ${m}$ as a function of reduced temperature $(T-T_{\rm C})/T_{\rm C}$ calculated for several different choices of external magnetic field ${H}_0$. Having NiPd alloy~\cite{Wei12} in mind, we use $UN(0) = 1.0001$, $A^{(0)}=10.0$, $b=60.0$, $c=20.0$, and assume $T_{\rm C}^{(0)}=100$K to reproduce $T_{\rm C}=20$K.

\begin{figure}[t] 
  \begin{center} 
    \includegraphics[width=6cm]{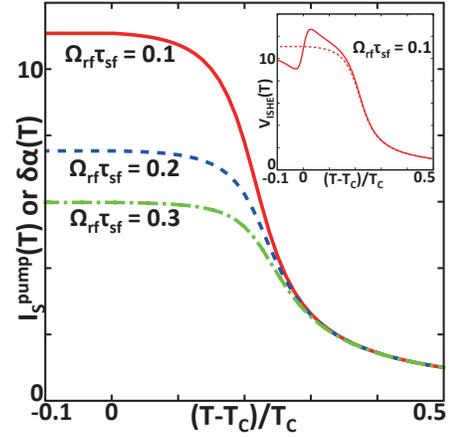}
  \end{center}
\caption{ 
(Color online) Pumped spin current $I_s^{\rm pump}$ [Eq.~(\ref{Eq:Is-total02b})] or additional Gilbert damping $\delta \alpha$ [Eq.~(\ref{Eq:Dalpha02})] at the resonance condition $\Omega_{\rm rf}=\gamma H_0$, calculated for a fluctuating SS (NiPd alloy) as a function of reduced temperature $(T-T_{\rm C})/T_{\rm C}$ with $\Omega_{\rm rf} \tau_{\rm sf}=0.1$ (solid line), $0.2$ (dashed line), and $0.3$ (dash-dotted line). All the data are normalized by their values at $T/T_{\rm C}=0.5$. Inset: Inverse spin Hall voltage used to electrically detect the enhanced spin pumping as a function of temperature, calculated using data from Ref.~\onlinecite{Wei12}. 
The dashed curve is given by the enhanced spin pumping predicted in this work, and the deviation from the dashed curve comes from the anomaly in the inverse spin Hall effect reported in Ref.~\onlinecite{Wei12}. For more details, see the main text. 
}
\label{fig5_Ohnuma}
\end{figure}

Now we investigate the enhanced spin pumping into the spin fluctuating SS. In Fig.~\ref{fig5_Ohnuma}, we show the temperature dependence of the pumped spin current $I_s^{\rm pump}$ for several choices of $\Omega_{\rm rf} \tau_{\rm sf}$. One can see that the pumped spin current is enhanced near $T_{\rm C}$. Because the pumped spin current $I_s^{\rm pump}$ and the additional Gilbert damping $\delta \alpha$ are intimately related through Eq.~(\ref{Eq:Is-Dalpha01}), this enhancement can be seen in the temperature dependence of the additional Gilbert damping as well. The enhancement is larger for a smaller value of $\Omega_{\rm rf} \tau_{\rm sf}$, which means that the enhancement is more visible in a material with a larger spin-orbit interaction. The case of NiPd alloy is estimated to be $\Omega_{\rm rf} \tau_{\rm sf} \sim 0.1$ using $\tau_{\rm sf} \approx 10^{-12}$~s, such that the enhancement can be as large as tenfold.

In experiments, the pumped spin current is detected electrically via the inverse spin Hall effect~\cite{Saitoh06}:
\begin{eqnarray}
  \bmE_{\rm ISHE} &=& \theta_{\rm SH} \rho \bmJ_s \times {\bm \sigma}, 
\end{eqnarray}
where $\bmE_{\rm ISHE}= -{\bm \nabla}V_{\rm ISHE} $ is the electric field induced by the inverse spin Hall effect, ${\bm \sigma} $ ($\parallel \hat{\bm z}$) is the spin polarization direction, $\theta_{\rm SH}$ and $\rho$ are respectively the spin Hall angle and the resistivity of the SS, and $\bmJ_s=(eI^{\rm pump}_s/A_{\rm int})\hat{\bm x}$ with the electronic charge $e$ is the spin-current density across the SI/SS interface having a contact area $A_{\rm int}$ (see Fig.~\ref{fig1_Ohnuma}). 

In the inset of Fig.~\ref{fig5_Ohnuma}, we plot the temperature dependence of the inverse spin Hall voltage calculated using the data from Ref.~\onlinecite{Wei12}. 
Note that the spin Hall angle in NiPd alloy near $T_{\rm C}$~\cite{Wei12} is decomposed into a temperature-independent background and the temperature-dependent component that reflects temperature dependence of the nonlinear susceptibility near $T_{\rm C}$~\cite{Wei12,Gu12}. The dashed curve is calculated using the temperature-independent component of the spin Hall angle, whereas the solid curve is calculated using the temperature-dependent spin Hall angle~\cite{Wei12}. In addition to a small structure coming from the anomaly in the inverse spin Hall effect, we see a clear enhancement of the inverse spin Hall voltage. Therefore, the predicted enhancement of $I_s^{\rm pump}$ can be detected electrically using the inverse spin Hall effect.

\section{Discussion and Conclusion \label{Sec:Conclusion}}

The main message of this paper is the theoretical prediction that the dc spin pumping can be largely enhanced if a fluctuating ferromagnet near $T_{\rm C}$ is used as the SS. Taking NiPd alloy as a prototype example, we have demonstrated that the enhancement can be as large as tenfold (Fig.~\ref{fig5_Ohnuma}). The underlying physics is as follows. As has been shown in Eq.~(\ref{Eq:Gs01}) the spin conductance across the SI/SS interface that characterizes the strength of the dc spin pumping is given by the square of the $s$-$d$ interaction $J_{\rm sd}$ at the SI/SS interface, multiplied by the imaginary part of the transverse spin susceptibility of the SS, ${\rm Im} \chi_\bmk (\omega)$. Because the latter quantity is known to be largely enhanced upon approaching $T_{\rm C}$, the interface spin conductance is increased near $T_{\rm C}$, resulting in a large enhancement of the dc spin pumping. 

Such kinds of many-body effects arising from critical spin fluctuations in the SS cannot be accounted for by the existing spin pumping theory~\cite{Tserkovnyak02} combined with the Landauer-type scattering approach~\cite{Tserkovnyak05}. In order to overcome this difficulty, we have developed a linear response formalism of the dc spin pumping, and calculated the fluctuation enhancement of the dc spin pumping by means of the SCR theory. Furthermore, a discussion is given in Appendix~\ref{Sec:Append03} on the opposite situation of the dc spin pumping {\it from} a fluctuating ferromagnet near $T_{\rm C}$ into a nonfluctuating SS. The temperature dependence of the dc spin pumping, which is expected from the knowledge on dynamic critical phenomena~\cite{Hohenberg77}, is calculated. 

Concerning the issue of spin injection into semiconductors, the spin pumping into GaAs and Si at room temperature has already been reported in Refs.~\onlinecite{Ando11} and \onlinecite{Ando12}. In the present context, it would be interesting to test the spin pumping {\it into} (Ga,Mn)As near its $T_{\rm C}$ in order to prove our prediction. Moreover, if a {\it fluctuating} room-temperature ferromagnetic semiconductor is discovered in the future, in spite of the smallness of its saturation magnetization and spin polarization at room temperature, we can achieve an efficient spin pumping into such a semiconductor by using the present scheme.

To summarize, we have developed a linear-response formalism of the dc spin pumping into a fluctuating ferromagnet near $T_{\rm C}$, and shown that the spin pumping can be largely enhanced owing to the fluctuation enhancement of the interface spin conductance. This effect may be used to construct an efficient spin current source using the dc spin pumping.

\acknowledgments 
We are grateful to M. Kimata, Y. Kajiwara, Y. Niimi, Y. Otani, and Y. Shiomi for valuable discussion. This work was financially supported by a Grant-in-Aid from MEXT, Japan, and a Fundamental Research Grant from CREST-JST, Japan.

\appendix

\section{Evaluation of the interface Green's function~\label{Sec:Append01}} 
In this appendix, we provide the procedure to evaluate the interface Green's function $C^<_{\bmk,\bmq}(t,t')$. Using the technique to calculate the contour-ordered Green's function, the diagram in Fig.~\ref{fig2_Ohnuma} is written as 
\begin{eqnarray}
  C^<_{\bmk,\bmq}(t,t') &=&
  \ui \frac{J_{\rm sd}(\bmk,\bmq)}{\hbar} \sqrt{\frac{S_0^3 N_{\rm SI}}{2 N_{\rm SS}}} 
  (\gamma h_{\rm rf})^2 \delta_{\bmq,{\bm 0}} \nonumber \\
  &\times& \int_{C} d \tau_{1} d\tau_{2} d\tau_{3} 
  \Big[ \chi_{\bmk}(t,\tau_{1}) \nonumber \\
    &\times& G^{(0)}_{0} (\tau_{2},\tau_{1})G^{(0)}_{0}(t',\tau_{3}) \Big]^< 
  e^{-\ui \Omega_{\rm rf}(\tau_{2}- \tau_{3})}, \qquad \label{Eq:A-Keldysh01} 
\end{eqnarray}
where $\tau_{1}, \tau_2, \tau_3$ are contour variables on the closed time path, $\chi_{\bmk}(t, t') =  -\ui \langle {\cal T}_C [s_{\bmk}(t) s_{-\bmk}^{+} (t')] \rangle$ is the contour-ordered transverse susceptibility of itinerant spins in the SS, $G^{(0)}_{0}(t,t')= -\ui \langle {\cal T}_C [b_{\bmq={\bm 0}}(t) b_{\bmq={\bm 0}}^\dag (t')] \rangle_0$ is the contour-ordered bare Green's function of uniform-mode magnon in the SI~\cite{Rammer86}, and $B^{\gtrless}$ means the greater/lesser part of Green's function $B$. It is convenient to introduce a convolution function 
\begin{eqnarray}
F_{\bmk}(t, \tau_{2}) &=& 
\int_{C} d\tau_{1}\chi_{\bmk}(t,\tau_{1}) G^{(0)}_{0}(\tau_{2},\tau_{1}). 
\end{eqnarray}

The integral over $\tau_2$ is evaluated in the following way. First, we deform the contour into the real-time contour~\cite{Rammer86}. In doing so, we use the fact that the exponentially oscillating factor has no singularity across the real-time axis, such that it can be dropped temporary in discussing the contour deformation. Thus, we have 
\begin{eqnarray}
   \int_{C}d\tau_{2} F_{\bmk}(t,\tau_{2}) 
   &=& \int_{-\infty}^{t}dt_{2} F^{>}_{\bmk}(t,t_{2})
   +\int_{t}^{-\infty}dt_{2} F^{<}_{\bmk}(t,t_{2})\nonumber \\
   &=& \int_{-\infty}^{\infty}dt_{2} F^{R}_{\bmk}(t,t_{2}), 
\label{Eq:A-Langreth01}
\end{eqnarray}
where $F^\gtrless$ and $F^R$ are explicitly given by 
\begin{eqnarray}
   F^{\gtrless}_{\bmk}(t,\tau_{2}) &=& \int_{-\infty}^{\infty} dt_{1} 
   \Big[\chi^{R}_{\bmk}(t,t_{1}) G^{(0)\lessgtr}_{0}(t_{1},t_{2}) \nonumber \\
   &&+ \chi^{\lessgtr}_{\bmk}(t,t_{1}) G^{(0)R}_{0}(t_{1},t_{2}) \Big] 
\label{Eq:A-X-R-L-G02}
\end{eqnarray}
and 
\begin{eqnarray}
   F^{R}_{\bmk}(t,\tau_{2}) &=& 
   \int_{-\infty}^{\infty}dt_{1} \chi^{R}_{\bmk}(t,t_{1}) G^{(0)A}_{0}(t_{2},t_{1}) 
\label{Eq:A-X-R-L-G01} 
\end{eqnarray}
with $B^R$ ($B^A$) being the retarded (advanced) part of a Green's function $B$. 

The integral over $\tau_3$ is performed in a similar way, giving 
\begin{eqnarray}
  \int_{C} d\tau_{3} G^{(0)}_{0}(t',\tau_{3}) &=& 
  \int_{-\infty}^{t}dt_{3} G^{(0)>}_{0}(t',t_{3}) \nonumber \\ 
  &&+ \int_{t}^{-\infty}dt_{3} G^{(0)<}_{0}(t',t_{3}) \nonumber \\ 
  &=& \int_{-\infty}^{\infty}dt_{3} G^{(0)R}_{0}(t',t_{3}). 
\label{Eq:A-Langreth02}
\end{eqnarray}
Substituting Eqs.~(\ref{Eq:A-Langreth01}) and (\ref{Eq:A-Langreth02}) into Eq.~(\ref{Eq:A-Keldysh01}), we obtain Eq.~(\ref{Eq:C-func01}) in Sec.~\ref{Sec:Form}. 

\section{Spin pumping from fluctuating ferromagnets \label{Sec:Append03}} 

In this appendix, we briefly discuss an issue of the dc spin pumping {\it from} a fluctuating ferromagnet near $T_{\rm C}$ into a nonfluctuating SS, such as a case of a EuO/Pt bilayer, by neglecting mode-mode coupling effects. Because we are interested in a temperature region near the Curie temperature of the SI, where the Landau-Lifshitz-Gilbert equation with fixed magnetization size is invalid, we begin with the following time-dependent Ginzburg-Landau type equation~\cite{Ma75}: 
\begin{eqnarray}
  \partial_t \widetilde{\bmS} &=& \gamma \bmH_{\rm eff} \times \widetilde{\bmS} 
  + \Gamma \frac{\bmH_{\rm eff}}{\mathfrak{h}_0} , 
  \label{Eq:TDGL01} 
\end{eqnarray}
where $\widetilde{\bmS}(\bmr,t)$ is the coarse-grained localized spin defined by 
\begin{eqnarray}
\widetilde{\bmS}(\bmr,t) &=& 
\frac{1}{\sqrt{N_{\rm SI}}} \sum_{q < 1/l_0} \bmS_\bmq e^{\ui \bmq \cdot \bmr}
\end{eqnarray}
with the momentum sum restricted to the wavelength being longer than a cutoff wavelength $l_0$~\cite{Ma75}, $\Gamma$ is the dissipative coefficient, and $\mathfrak{h}_0$ is a unit of magnetic field defined below Eq.~(\ref{Eq:Heff01}). If we were concerned about a spin-conserving system, the dissipative coefficient $\Gamma$ would be given by a spin diffusion process and expressed as $\Gamma = - {\cal D} \nabla^2$ with the spin diffusion coefficient ${\cal D}$~\cite{Hertz76}. However, because we are dealing with a spin nonconserving system with spin-orbit interaction, we set $\Gamma$ to be a constant $\Gamma_0$. 

The effective magnetic field $\bmH_{\rm eff}$ is given by 
\begin{equation}
\bmH_{\rm eff} = \bmH_0 + \bmh_{\rm rf} - \frac{{\cal J}_{\rm sd}}{\gamma \hbar} \bms 
- \frac{v_0}{\gamma \hbar} \frac{\delta F_{\rm GL}}{\delta \widetilde{\bmS}}, 
\end{equation}
where $\bmH_0$ is the uniform external magnetic field, 
$  \bmh_{\rm rf} = h_{\rm rf} \cos (\Omega_{\rm rf} t) {\bm \hat{\bmx}}- h_{\rm rf} \sin (\Omega_{\rm rf} t) {\bm \hat{\bmy}}$ is the oscillating magnetic field, 
${\cal J}_{\rm sd}(\bmr) \bms/ \gamma \hbar= (J_{\rm sd} \bms/\gamma \hbar) \sum_{\bmr_0 \in {\rm SI}/{\rm SS}} \delta_{\bmr_0, \bmr}$ 
describes the effect of spin accumulation $\bms$ with $J_{\rm sd}$ being the $s$-$d$ interaction at the SI/SS interface, and $v_0= l^3_0$ is the volume of a coarse-grained block spin volume. The free energy $F_{\rm GL}$ in the last term is given in the Ginzburg-Landau form~\cite{Chaikin-text}: 
\begin{eqnarray}
  F_{\rm GL} &=& \varepsilon_0 
  \int d^3 r \big( \frac{a_{\rm GL}}{2} \widetilde{S}^2+ \frac{b_{\rm GL}}{4} \widetilde{S}^4 \big), 
\end{eqnarray}
where $\varepsilon_0$ is the magnetic energy density, $a_{\rm GL} = (T-T_{\rm C})/T_{\rm C}$ measures the distance from the Curie temperature $T_{\rm C}$, $b_{\rm GL}$ is the quartic term coefficient, and the gradient term is discarded because it is sufficient to consider only the uniform mode dynamics for the present discussion on the FMR-driven spin pumping. If necessary, these coefficients can be determined from material parameters as 
\begin{eqnarray}
  \varepsilon_0 &=& \frac{ \Delta C}
             {\frac{d}{dT} \left[\frac{M^2_s(T)}{M^2_s(0)}\right]_{T_{\rm C}} }, \\
 b_{\rm GL} &=& \frac{1}{T_{\rm C} \, \frac{d}{dT} 
   \left[\frac{M^2_s(T)}{M^2_s(0)}\right]_{T_{\rm C}}},
\end{eqnarray}
where $\Delta C$ is the specific heat jump per unit volume in the limit of the mean-field approximation and $M_s(T)$ is the saturation magnetization at a temperature $T$. Under a finite uniform magnetic field $\bmH_0$ to realize the magnetic resonance, the equilibrium localized spin $\widetilde{S}_{\rm eq}$ is determined by the equation 
\begin{eqnarray}
  H_0 &=& \mathfrak{h}_0 \big( a_{\rm GL} \widetilde{S}_{\rm eq} 
  + b_{\rm GL} \widetilde{S}_{\rm eq} ^3 \big), 
  \label{Eq:Heff01} 
\end{eqnarray}
where $\mathfrak{h}_0=\varepsilon_0 v_0/\gamma \hbar$ gives the unit of a magnetic field (a very crude estimate for EuO gives $\mathfrak{h}_0$ to be of the order of one tesla~\cite{Borukhovich76}). This equation is derived from the condition $\bmH_{\rm eff}= {\bm 0}$ in the absence of $\bmh_{\rm rf}$ and $\bms$. In the following, we measure the strength of the magnetic field in the unit of $\mathfrak{h}_0$ and introduce $\widetilde{H}_0= H_0/\mathfrak{h}_0$, and the size of the localized spin $\widetilde{S}$ is measured with respect to its zero temperature value. 

\begin{figure}[t] 
  \begin{center}
    \includegraphics[width=9cm]{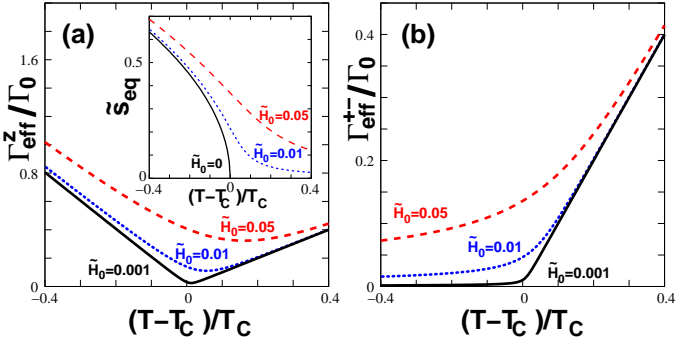}
  \end{center}
\caption{ 
(Color online) (a) Effective {\it longitudinal} damping coefficient $\Gamma^{z}_{\rm eff}$ as a function of reduced temperature. Inset: Temperature dependence of equilibrium spin $\widetilde{S}_{\rm eq}$. (b) Effective {\it transverse} damping coefficient $\Gamma^{+-}_{\rm eff}$ as a function of reduced temperature. In both figures, $b_{\rm GL}=1.0$ is used. 
}
\label{fig_Gam01}
\end{figure}

Noticing that $\delta F_{\rm GL}/ \delta \widetilde{\bmS} $ is parallel to $\widetilde{\bmS}$ and using the value of $\widetilde{S}_{\rm eq}$, Eq.~(\ref{Eq:TDGL01}) can be rewritten as 
\begin{eqnarray}
  \partial_t \widetilde{\bmS} &=& \gamma (\bmH_0+ \bmh_{\rm rf} 
  - \frac{J_{\rm sd}}{\gamma \hbar} \bms ) \times \widetilde{\bmS} 
  - \overleftrightarrow{\Gamma}_{\rm eff} (\widetilde{\bmS} - \widetilde{\bmS}_{\rm eq} ), 
  \qquad 
 \label{Eq:TDGL02}
\end{eqnarray}
where we have defined the effective damping tensor 
$\overleftrightarrow{\Gamma}_{\rm eff}= {\rm diag}(\Gamma^{+-}_{\rm eff}, \Gamma^{+-}_{\rm eff},\Gamma^z_{\rm eff}) $ with $\Gamma^{+-}_{\rm eff}= \Gamma_0 (a_{\rm GL}+b_{\rm GL} \widetilde{S}^2_{\rm eq})$ and $\Gamma^z_{\rm eff}= \Gamma_0 (a_{\rm GL}+ 3b_{\rm GL} \widetilde{S}^2_{\rm eq})$. 
We combine Eq.~(\ref{Eq:TDGL02}) with the following Bloch equation for $\bms$: 
\begin{eqnarray}
  \partial_t \bms &=& - \frac{J_{\rm sd}}{\hbar} \widetilde{\bmS} \times \bms 
  + D \nabla^2 \bms 
  - \frac{1}{\tau_{\rm sf}} (\bms - s_0 \widetilde{\bmS}), 
  \label{Eq:Bloch01}
\end{eqnarray}
where $D$ is the spin diffusion coefficient of the SS, $\tau_{\rm sf}$ is the spin-flip relaxation time in the SS, and $s_0=\chi_{\rm P} J_{\rm sd}$ is the local equilibrium spin density with $\chi_{\rm P}$ being the Pauli paramagnetic susceptibility of the SS. 

Starting from Eqs.~(\ref{Eq:TDGL02}) and (\ref{Eq:Bloch01}), we calculate the spin current pumped into the nonmagnetic SS. We first define the pumped spin current $I_s^{\rm pump}$ as the rate of change in the itinerant spin density in the nonmagnetic SS as $I_s^{\rm pump}=\langle \partial_t s^z(t) \rangle$. Then, performing the perturbative approach in the Bloch equation (\ref{Eq:Bloch01}) with respect to $J_{\rm sd}$, we obtain 
\begin{eqnarray}
  I_s^{\rm pump}(t) &=& \frac{J_{\rm sd}}{\hbar N_{\rm SS}} \sum_{\bmk} {\rm Im} \langle S_{\bmq={\bm 0}}^-(t) s_{\bmk}^+(t) \rangle, 
  \label{Eq:IsGL01}
\end{eqnarray}
where $S^\pm = S^x \pm \ui S^y$ and $s^\pm = s^x \pm \ui s^y$. 
Introducing the Fourier representation $f(t)=\int \frac{d \omega}{2 \pi} f(\omega) e^{-\ui \omega t}$, we obtain 
\begin{eqnarray}
  I^{\rm pump}_s &=& \frac{J_{\rm sd}}{\hbar N_{\rm SS}}  
  \sum_\bmk
      {\rm Im} \langle S_{\bmq=0}^-(\Omega_{\rm rf}) s_\bmk^+(-\Omega_{\rm rf}) \rangle 
       \label{Eq:IsGL02} 
\end{eqnarray}

To evaluate the right hand side of Eq.~(\ref{Eq:IsGL02}), the transverse components of Eqs.~(\ref{Eq:TDGL02}) and (\ref{Eq:Bloch01}) are linearized with respect to $S^\pm$ and $s^\pm$. Then, to the lowest order in $J_{\rm sd}$, we obtain 
\begin{eqnarray} 
  S_{\bmq={\bm 0}}^- (\Omega_{\rm rf}) &=& - G_0(\Omega_{\rm rf}) 
  \gamma h^-_{\rm rf} 
  \label{Eq:S-01}
\end{eqnarray}
and 
\begin{eqnarray}
  s_\bmk^+(\Omega_{\rm rf}) &=& - s_0 \widetilde{\chi}_\bmk (\Omega_{\rm rf}) G_0^*(-\Omega_{\rm rf}) 
  \gamma h_{\rm rf}^+, 
  \label{Eq:s+01}
\end{eqnarray}
where $h_{\rm rf}^\pm= h_{\rm rf}^x \pm \ui h_{\rm rf}^y$. Here, we have introduced the (normalized) paramagnetic susceptibility 
\begin{eqnarray}
\widetilde{\chi}_\bmk (\omega) &=& \frac{1}{1+ \lambda^2 k^2 - \ui \omega \tau_{\rm sf}} 
\end{eqnarray}
as well as the ferromagnetic susceptibility 
\begin{eqnarray}
  G_0(\omega) &=& \frac{\widetilde{S}_{\rm eq}}
  {\omega- \gamma H_0 + \ui \Gamma^{+-}_{\rm eff} }, 
\end{eqnarray}
where $\lambda=\sqrt{D \tau_{\rm sf}}$ is the spin-diffusion length. Note that the critical slowing down manifests itself in the shrinking of the damping term $\Gamma^{+-}_{\rm eff} \ll \Gamma_0$ on approaching the Curie temperature $T_{\rm C}$. Substituting the above equations into Eq.~(\ref{Eq:IsGL02}), the spin current pumped into the SS can be expressed as 
\begin{eqnarray}
  I^{\rm pump}_s &=& - \frac{J^2_{\rm sd}}{\hbar N_{\rm SS}} 
  \sum_\bmk 
      {\rm Im} {\chi}_\bmk (\Omega_{\rm rf}) | G_0 (\Omega_{\rm rf})|^2
      (\gamma h_{\rm rf})^2 \qquad \\
      &=& \mathfrak{g}_s \frac{\Omega_{\rm rf} (\gamma h_{\rm rf})^2}
      {(\Omega_{\rm rf}-\gamma H_0)^2 + (\Gamma^{+-}_{\rm eff})^2}, 
      \label{Eq:Is-SpinPump01}
\end{eqnarray}
where $\mathfrak{g}_s = \frac{(J_{\rm sd}\widetilde{S}_{\rm eq})^2}{\hbar^2 N_{\rm SS}} \sum_\bmk \frac{1}{\Omega_{\rm rf}} {\rm Im} \chi_\bmk (\Omega_{\rm rf})$, and we have defined the dynamical transverse susceptibility $\chi_\bmk (\omega)= \chi_{\rm P} \widetilde{\chi}_{\bmk}(\omega)$.

\begin{figure}[t] 
  \begin{center}
    \includegraphics[width=5cm]{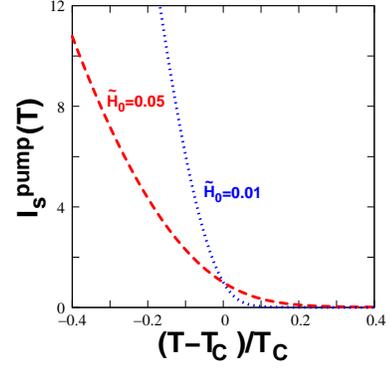}
  \end{center}
\caption{ 
(Color online) Temperature dependence of pumped spin current near $T_{\rm C}$. The data is normalized by its value at $T=T_{\rm C}$, and $b_{\rm GL}=1.0$ is used. 
}
\label{fig_ESRpump}
\end{figure}

Figure~\ref{fig_Gam01}(a) shows the effective {\it longitudinal} damping coefficient $\Gamma^{z}_{\rm eff}$ introduced below Eq.~(\ref{Eq:TDGL02}) as a function of temperature for several different values of an external magnetic field $\widetilde{H}_0$. Temperature dependence of the equilibrium spin value is also plotted in the inset. Because the longitudinal spin dynamics is strongly related to the spin diffusion~\cite{Ma75,Kawasaki67}, the critical slowing down (shrinking of $\Gamma^{z}_{\rm eff}$) in the limit of an infinitesimally small external field appears almost symmetrically for $T>T_{\rm C}$ and $T<T_{\rm C}$. By contrast, the effective {\it transverse} damping coefficient $\Gamma^{+-}_{\rm eff}$ does not show such a symmetric critical slowing down across $T_{\rm C}$. 

Figure~\ref{fig_Gam01}(b) shows the effective transverse damping coefficient $\Gamma^{+-}_{\rm eff}$ as a function of temperature for several different values of an external magnetic field $\widetilde{H}_0$. Upon lowering the temperature across $T_{\rm C}$ the transverse damping coefficient $\Gamma^{+-}_{\rm eff}$ keeps decreasing, in contrast to the behavior of $\Gamma^{z}_{\rm eff}$. This calculated behavior is consistent with experimental results for iron~\cite{Bhagat72} and yttrium iron garnet~\cite{Berzhanskii89}, once recalling that the transverse damping coefficient $\Gamma^{+-}_{\rm eff}$ is proportional to the FMR linewidth near $T_{\rm C}$. Note that the transverse damping coefficient $\Gamma^{+-}_{\rm eff}$ is also responsible for the dc spin pumping given by Eq.~(\ref{Eq:Is-SpinPump01}).

In Fig.~\ref{fig_ESRpump}, the pumped spin current calculated from Eq.~(\ref{Eq:Is-SpinPump01}) is shown as a function of temperature. Upon lowering the temperature, the pumped spin current is largely enhanced owing to the shrinking of the linewidth $\Gamma^{+-}_{\rm eff}$ and the increase of the interface spin conductance $\mathfrak{g}_s \propto \widetilde{S}^2_{\rm eq}$. The overall temperature dependence looks consistent with an experimental result of the spin pumping from (Ga,Mn)As into {\it p}-type GaAs reported in Ref.~\onlinecite{Chen13}. Note that the dc spin pumping appears even in the paramagnetic region above $T_{\rm C}$ under a condition of sizable external magnetic field to obtain the magnetic resonance, which means that the spin pumping can be driven by electron paramagnetic resonance (EPR) above $T_{\rm C}$.


\end{document}